\documentclass[10pt,a4paper,twoside]{article}

\usepackage{amssymb,amsmath,multirow, graphics,color}
\usepackage{comment}
\newtheorem{theorem}{Theorem}[section]

\usepackage[dvips]{graphicx}
\usepackage{epsfig}
\usepackage{epstopdf}
\usepackage{authblk}

    \makeatletter
    \let\@fnsymbol\@arabic
    \makeatother

\begin{document}

	\title{Asymptotic Efficiency of Goodness-of-fit Tests Based on  Too-Lin  Characterization  }
	\author{ Bojana Milo\v sevi\'c \footnote{ bojana@matf.bg.ac.rs} \\ \bigskip
		{\small Faculty of Mathematics, University of Belgrade, Belgrade, Serbia} }

	\date{}
	
	\maketitle

	\begin{abstract}
		{In this paper a new class of  uniformity tests  is proposed. It is shown that those tests are applicable to the cases of any simple null hypothesis as well as for  the composite null hypothesis of rectangular distributions on arbitrary support.  The asymptotic properties of  test statistics  are examined. The tests are compared with some standard and some recent uniformity tests. For each test the  Bahadur efficiencies  against some common local alternatives are calculated. A class of locally optimal alternatives is found for each proposed test. The power study is also provided. Some  applications in time series analysis are  presented.}
	\end{abstract}

	{\small \text{ keywords:} testing uniformity, moments of order statistics,\\ Bahadur efficiency, $U$-statistics, conditional duration models.
		
		\text{MSC(2010):} 60F10,\ 62G10, \ 62G20,\ 62G30}
	
	\section{Introduction}
	The uniform distribution is one of the most used distribution in statistical modeling and computer science.
	Therefore ensuring that the data come from uniform distribution is of  huge importance. Moreover, testing that the data come from a particular distribution can be easily reduced to testing uniformity. More about such  goodness-of-fit techniques   can be found in \cite{agostino} and \cite{dzhaparidze1982probability}.
	
	In recent times, the tests based on some characteristic  property that distribution possesses  have become very popular. Many  different types of characterizations can be found e.g in \cite{galambos}.   Some characterizations of the uniform distribution can be found e.g in \cite{anand}, \cite{goria}, \cite{volkmer}.
	The first uniformity test  based on a characterization that involves moments, was proposed  in \cite{hashimoto}. Their test  was based on Papathanasiou's characterization on maximal covariance  between the minimum  and  the  maximum of   a  sample of  the  size two, presented in \cite{papathanasiou}. Other tests based on characterizations via moments have been considered in,  among others,  \cite{morris2000}, \cite{morris2004}.
	
	One way to compare tests is to calculate their asymptotic  efficiencies. In  the  case of non-normal limiting distribution, the Bahadur approach to efficiency is suitable (see \cite{Bahadur} and \cite{nikitinKnjiga}). Among recent papers, it has been considered  in e.g. \cite{Tenreiro}, \cite{NikVol}, \cite{graneTchirina}, \cite{jovanovic},  \cite{samostalni1}, \cite{obradovic},   \cite{bojana}, \cite{laplace}, \cite{2dKS}.

	In this paper we propose   new  uniformity tests based on  the  characterization from \cite{tooLin},  that involves some moments of order statistics.  We examine the asymptotic properties of the test statistics. This characterization  has already been used  in \cite{morris2000}, however the nature of their test is completely different.
	
	
	The paper is organized as follows. In Section \ref{sec:characterizations} we present the characterization and  a  class of test statistics based on it.
	Next, we give   a  brief introduction to Bahadur theory. 
	Using  the  Bahadur efficiency we compare our test with Hashimoto-Shirahata test based on  a  maximal covariance characterization (see \cite{hashimoto}),  Fortiana-Gran\'e  test based on maximum correlations (see \cite{graneTchirina}), as well as some standard goodness-of-fit tests.  Additionally,  for each proposed test we find some classes of locally optimal alternatives. Next we compare  the  Bahadur efficiencies of presented tests in  the  case of testing   the  null hypothesis  of standard normal, standard logistic and standard Cauchy distribution against location alternatives. Finally, we adapt our tests for testing "rectangularity" on  an  unknown support. 
	In Section \ref{sec:power} we perform   a  power study and present some applications in time series analysis.
	\section{Characterization and Test Statistics}\label{sec:characterizations}
	
	In \cite{tooLin}  the following characterization of  the  uniform distribution is proved.
	
	\begin{theorem} Let the  kth order statistic from the  i.i.d.  sample of size $m$,  $X_{(k),m}$  has finite second moment,  $EX_{(k),m}^2<\infty$ for some pair $(k, m)$.
		Then the equality
		\begin{equation}
			EX^2_{(k),m}-\frac{2k}{m+1}EX_{(k+1),m+1}+\frac{k(k+1)}{(m+1)(m+2)}=0
		\end{equation}
		holds if and only if $F(x)= x$ on $(0, 1)$.
	\end{theorem}
	
	
	Denote $X_{(a),X_1,...,X_n}$ the $a$th order statistic of the i.i.d.
	sample $X_1,...,X_n$.   We test the null hypothesis $H_0: (F(x)=x,\;x\in(0,1))$.   In  view of  the characterization for $k=1$, we
	propose the following class of test statistics.
	
	\begin{align*}
		T^{(m)}_n&=\frac{1}{\binom{n}{m+1}}\sum\limits_{i_1<\cdots<i_{m+1}}\Big(\frac{1}{(m+1)!}\sum\limits_{\pi\in \Pi(m+1)}\!\!\!
		X_{(1),X_{i_{\pi(1)}},...,X_{i_{\pi(m)}}}^2\\&-\frac{2}{m+1}
		X_{(2),X_{i_1},...,X_{i_{m+1}}}+\frac{2}{(m+1)(m+2)}\Big),
	\end{align*}
	where $\Pi(m)$ is the set of all one-to-one mappings $\pi:\{1,....,m\}\mapsto \{1,....,m\}.$ 
	
	We consider large  absolute values of test statistic to be significant.
	
	Notice that these statistics are $U$-statistics with symmetric kernels
	
	\begin{align*}
		\Phi_m(X_1,..,X_{m+1})&=\frac{1}{(m+1)!}\sum\limits_{\pi\in \Pi(m+1)}\!\!\!X_{(1),X_{\pi(1)},...,X_{\pi(m)}}^2
		-\frac{2}{m+1}X_{(2),X_{1},...,X_{{m+1}}}\\&+\frac{2}{(m+1)(m+2)},
	\end{align*}
	where $\Pi(m)$ is the set of all permutations of numbers $1,2,...,m$.
	
	The first projections of  the  kernels $ \Phi_m(X_1,..,X_{m+1})$ on $X_{m+1}$ under $H_0$ are
	\begin{align*}
		\phi_m(s)&= E(\Phi_m(X_1,..,X_{m+1})|X_{m+1}=s)=\frac{m}{m+1}E(X^2_{(1),s,X_1,...,X_{m-1}})\\&+\frac{1}{m+1}E(X^2_{(1),X_1,...,X_{m}})-\frac{2}{m+1}E(X_{(2),s,X_{1},...,X_{m}})+\frac{2}{(m+1)(m+2)}.
	\end{align*}
	
	We have
	
	\begin{align*}
		E(X^2_{(1),s,X_1,...,X_{m-1}})&=E(X^2_{(1),X_1,...,X_{m-1}}I\{X_{(1),X_1,...,X_{m-1}}<s\})\\&+s^2P\{X_{(1),X_1,...,X_{m-1}}>s\}
		\\&=\frac{2-2 (1-s)^m (m s+1)}{m(m+1)}.\\
	\end{align*}
	Similarly,
	\begin{align*}
		E(X^2_{(1),X_1,...,X_{m}})&=\int_{0}^1y^2\cdot m(1-y)^{m-1}dy=\frac{2}{(m+1)(m+2)},\\
		E(X_{(2),s,X_1,...,X_{m}})&= E(X_{(1),X_1,...,X_{m}}I\{s<X_{(1),X_1,...,X_{m}}\})\\&+sP\{X_{(1),X_1,...,X_{m}}<s<X_{(2),X_1,...,X_{m}}\}\\&+E(X_{(2),X_1,...,X_{m}}I\{s>X_{(2),X_1,...,X_{m}}\})\\&=\int_{s}^1y\cdot m(1-y)^{m-1}dy\\&+s\cdot ms(1-s)^{m-1}\\&+\int_{0}^sy\cdot m(m-1)y(1-y)^{m-2}dy\\
		&=\frac{2-(1-s)^m (m s+1)}{m+1}.
	\end{align*}
	Therefore, the first projections are  equal to zero under  the  null hypothesis. Hence, the statistics $T^{(m)}_n$ are degenerate for every $m$.

	
	The second projections of $\Phi_m(X_1,..,X_{m+1})$ on $(X_{m},X_{m+1})$ under  the   null hypothesis  are
	
	\begin{align*}
		\phi^{\ast}_m(s,t)&=E(\Phi_m(X_1,..,X_{m+1})|X_{m}=s,X_{m+1}=t)\\&=\frac{m-1}{m+1}E(X^2_{(1),s,t,X_{1},...,X_{m-2}})
		+\frac{1}{m+1}E(X^2_{(1),s,X_{1},...,X_{m-1}})\\&+\frac{1}{m+1}E(X^2_{(1),t,X_{1},...,X_{m-1}})
		-\frac{2}{m+1}E(X_{(2),s,t,X_{1},...,X_{m-1}})\\&+\frac{2}{(m+1)(m+2)}.
	\end{align*}
	Using the same reasoning as before, after some calculations, we obtain
	\begin{align*}
		\phi^{\ast}_m(s,t)&=-\frac{2}{m (1 + m)^2 (2 + m)}
		\Big(-2 + 2(1 - t)^m +
		m^2 \big(-(1 - s)^{m+1} \\& + (1 - t)^m t\big) +
		m \big(-2 (1 - s)^{m+1}  + (1 - t)^m +
		2 (1 - t)^m t\big)\Big)\\& +\frac{2}{m (1 + m) }I\{s<t\} ( (1 - t)^m-(1 - s)^m ).
	\end{align*}
	Obviously,  $\phi^{\ast}_m$   is not equal to zero for any choice of $m$. Therefore we conclude that the
	kernels of our test statistics are weakly degenerate. 
	
	Using   Theorem 4.4.1  from  \cite{Kor} for weakly degenerate $U$-statistics we have
	
	\begin{equation*}
		nT^{(m)}_n\overset{d}{\rightarrow}\binom{m+1}{2}\sum\limits_{i=1}^{\infty}\nu^{(m)}_i(\tau^2_i-1),
	\end{equation*}
	where ${\nu^{(m)}_i}$ are the eigenvalues of the integral operator $S$
	defined by
	\begin{equation}\label{operatorS}
		Sf(t)=\int\limits_{0}^1\phi^{\ast}_m(s,t)f(s)ds.
	\end{equation}
	
	Thus we have to solve the following integral equation
	\begin{equation}\label{integral equation}
		\nu^{(m)} f(t)=\int\limits_{0}^1\phi^{\ast}_m(s,t)f(s)ds,
	\end{equation}
	with constraint $\int\limits_{0}^1f(s)ds=0$.
	
	Denote $y(t)=\int_0^tf(s)ds$. After  the  differentiation, the expression \eqref{integral equation} becomes
	\begin{equation*}
		\nu^{(m)}  y''(t)=-2\frac{(1-t)^{m-1}}{m+1}y(t),\;\;\;y(0)=0,\;\; y(1)=0.
	\end{equation*}
	After  the change of variables   $s=1-t$, $\lambda^{(m)} =\frac{1}{\nu^{(m)} },$ we obtain  the  following
	boundary problem
	\begin{equation}\label{jednacinaLambda}
		y''(s)+\frac{2}{m+1}\lambda^{(m)}  y(s)s^{m-1}=0,\;y(0)=1,\;y(1)=1.
	\end{equation}

	Using  the   result from \cite[2.162, p.440]{Kamke} for $a=0$, $c=0$ and
	$b\neq 0$ we obtain  the  solution as a linear combination of Bessel's
	functions of the first kind
	
	\begin{align*}
		y(s)=\sqrt{s}\Big(C_1J_{-\frac{1}{m+1}}\Big((\frac{2}{m+1})^{\frac{3}{2}}\sqrt{\lambda^{(m)}}s^{\frac{m+1}{2}}\Big)+C_1J_{\frac{1}{m+1}}\Big((\frac{2}{m+1})^{\frac{3}{2}}\sqrt{\lambda^{(m)}}s^{\frac{m+1}{2}}\Big)\Big).
	\end{align*}
	
	From  the  boundary conditions we have

	\begin{equation}\label{jednacinaBesel}
		\begin{aligned}
			&0=C_1,\\
			&0=J_{-\frac{1}{m+1}}\Big(\frac{2\sqrt{2\lambda^{(m)}}}{(1+m)^{\frac{3}{2}}}\Big)C_1+J_{\frac{1}{m+1}}\Big(\frac{2\sqrt{2\lambda^{(m)}}}{(1+m)^{\frac{3}{2}}}\Big)C_2.
		\end{aligned}
	\end{equation}

	Hence,   the initial problem transforms into  the  problem of finding
	zeros of Bessel's function $J_{\frac{1}{m+1}}(x)$, which can be found
	for every $m\geq 1$,  numerically.

	Notice that for $m=1$ it is   Sturm-Liouville problem
	\begin{equation}\label{jednacinam1}
		y''(t)+\lambda^{(1)} y(t)=0,\; y(0)=0,\; y(1)=0.
	\end{equation}
	whose solution is well known (see e.g. \cite{metron}).

	\section{Local Bahadur Efficiency}\label{sec:bahadur}
	
	We choose Bahadur asymptotic efficiency as a measure of the quality
	of tests. One of the reasons is that the asymptotic distributions of
	our test statistics are not normal. The Bahadur efficiency   can be
	expressed as the ratio of the Bahadur exact slope, a function
	describing the rate of exponential decrease for the attained level
	of significance  under the alternative, and the double
	Kullback-Leibler distance between the null and the alternative
	distribution. We present  a  brief review of the theory, for more  about
	this topic  we refer to \cite{Bahadur}, \cite{nikitinKnjiga}.

	According to Bahadur's theory, the exact slopes  can be found   in the
	following way. Suppose that under alternative \begin{equation*}
		T_n\overset{P_\theta}{\rightarrow}b(\theta).
	\end{equation*}
	Also suppose that the large deviation limit
	\begin{equation}\label{ldf}
		\lim_{n\to\infty} n^{-1} \ln
		P_{H_0} \left( T_n \ge t  \right)  = - f(t)
	\end{equation}
	exists for any $t$ in an open interval $I,$ on which $f$ is
	continuous and $\{b(\theta), \: \theta > 0\}\subset I$. Then the
	Bahadur exact slope is     
	\begin{equation}\label{slope}
		c_T(\theta) = 2f(b(\theta)).
	\end{equation}
	
	The Bahadur-Raghavachari
	inequality
	\begin{equation}
		\label{Ragav} c_T(\theta) \leq 2 K(\theta),\, \theta > 0,
	\end{equation}
	where $K(\theta)$ is the Kullback-Leibler distance between the
	alternative $H_1$ and the null hypothesis $H_0$, leads to  a  natural
	definition of  the  Bahadur efficiency,  as the ratio of the  $c_T(\theta)$ and $2K(\theta).$    Since it is very important for a
	test to distinguish  close alternatives from   the  null distribution, we
	consider  the   local Bahadur efficiency, defined as
	\begin{equation}\label{localBahadurEf}
		e^B (T) = \lim_{\theta \to 0} \frac{c_T(\theta)}{2K(\theta)}.
	\end{equation}
	
	Let $\mathcal{G}=\{G(x;\theta): \theta\in[0,\theta^\star]\}$,  for $\theta^\star>0$,  be the class of absolutely continuous distribution functions with densities $g(x;\theta)$ satisfying the following conditions:
	\begin{itemize}
		\item  $G(x;\theta)$ is the d.f. of uniform $\mathcal{U}[0,1]$ random variable if and only if $\theta=0$;
		\item $g(x;\theta)$ is  three times  continuously differentiable along $\theta$ in some neighbourhood of zero;
		\item the partial derivatives of $g(x;\theta)$ along $\theta$, $g'_{\theta}(x;\theta)$, $g''_{\theta \theta}(x;\theta)$, and $g'''_{\theta \theta \theta}(x;\theta)$, are absolutely integrable for $\theta$ in some neighbourhood of zero.
	\end{itemize}
	Denote $h(x)=g'_{\theta}(x,0)$.

	It can be shown that the double Kullback-Leibler distance between the
	null    distribution and the close alternative can be expressed as
	\begin{equation}\label{KLFiser}
		2K(\theta)= I(g)\theta^2 + o(\theta^2),\;\;\theta\rightarrow 0,
	\end{equation}
	where  $I(g)\in(0,\infty)$ is the Fisher information function
	\begin{equation*}
		I(g)=\int\limits_0^1 \frac{h^2(x)}{g(x,0)}dx.
	\end{equation*}
	
	\medskip
	
	Note that the condition $\theta>0$ is no loss of generality. Any close distribution can be reparametrized such that $\theta \in[0,\theta^\star]$.

	\subsection{Statistic $T^{(m)}_n$}
	
	In the following theorem we give the expression for the exact local Bahadur slope of statistic  $T^{(m)}_n$.  
	
	\begin{theorem} \label{slopeNasa} Let $X_1,X_2...,X_n$ be an i.i.d. sample from an absolutely continuous  alternative distribution from $\mathcal{G}$. Then the local Bahadur slope of statistic $T^{(m)}_n$ is
		\begin{equation*}
			c_{T}(\theta)=\lambda^{(m)}_1\Big|\int_{0}^1\int_{0}^1h(s)h(t)\phi^{\ast}_m(s,t)dsdt\Big|\cdot\theta^2+o(\theta^2),\;\;\;\theta\to 0,
		\end{equation*}
		where $ \lambda^{(m)}_1=\frac{1}{\nu^{(m)}_1}$ , and $\nu^{(m)}_1$ is the largest eigenvalue of integral operator \eqref{operatorS}.
	\end{theorem}
	
	\textbf{Proof.}  Since the large absolute values are significant, we need the large deviation function and the limit in probability of $T^{(m)}_n$. 
	Taking into account that the support of the limiting distribution of  $T^{(m)}_n$ is semi-infinite, with its left end being finite, the large deviation function of  $|T^{(m)}_n|$ coincides with the one of the statistic  $T^{(m)}_n$.  Applying the result from \cite{nikiponi}, for weakly degenerate $U$-statistics, and the same reasoning as in  the proof \cite[Th. 4]{metron}, we complete the proof. \hfill{$\Box$}.

	For $m=1$, the equation  \eqref{jednacinam1} has  the  minimal solution $\lambda^{(1)}_1=\pi^2$. It is
	interesting to note that  Cramer-von Mises statistic has the same
	kernel (see e.g. \cite{metron}).  Therefore they  are asymptotically equivalent in the Bahadur sense. 
	
	In case of  $m=2$ we numerically obtain that the minimal solution of \eqref{jednacinaBesel} 
	is  $\lambda^{(2)}_1=28.4344$.

	\subsection{Competitor tests}
	We compare our tests with the following tests:
	\begin{itemize}
		\item Kolmogorov-Smirnov test with test statistic
		\begin{equation}\label{kolmogorov}
			D_n=\sup\limits_{t\in(0,1)}|F_n(t)-t|;
		\end{equation}
		\item Anderson-Darling test with test statistic
		\begin{equation}\label{andersond}
			A_n^2=\int\limits_{0}^1\frac{(F_n(t)-t)^2}{t(1-t)}dt;
		\end{equation}
		\item Cramer-von Mises test with test statistic
		\begin{equation}
			\omega_n^2=\int\limits_{0}^1(F_n(t)-t)^2dt;
		\end{equation}
		\item the test based on maximum correlations (see \cite{fortianaGrane}) with test statistic
		\begin{equation}\label{maxcor}
			Q^{c}_n=|Q_n-1|=|\frac{6}{n^2}\sum\limits_{i=1}^n(2 i-n-1)X_{(i)}-1|;
		\end{equation}
		\item   the  test  based on   the  maximal covariance characterization (see \cite{hashimoto}) with test statistic
		\begin{equation}\label{hashimoto}
			C_n=\binom{n}{4}^{-1}\sum_{i<j<k<l}h(X_i,X_j,X_k,X_l);
		\end{equation}
		where
		\begin{align*}
			&h(X_i,X_j,X_k,X_l)=\frac{1}{36}\Big((X_i-X_j)^2+(X_i-X_k)^2+(X_i-X_l)^2\\&+(X_j-X_k)^2+(X_j-X_l)^2+(X_k-X_l)^2\Big)
			\\&-\frac{1}{6}\Big((\max(X_i,X_j)-\max(X_k,X_l))(\min(X_i,X_j)-\min(X_k,X_l)\\&+(\max(X_i,X_k)-\max(X_j,X_l))(\min(X_i,X_k)-\min(X_j,X_l)
			\\&+(\max(X_i,X_l)-\max(X_j,X_k))(\min(X_i,X_l)-\min(X_j,X_k))\Big).
		\end{align*}
	\end{itemize}

	We calculate  the  local Bahadur efficiencies of  the  proposed tests and the
	competitor ones against the following alternatives:
	\begin{itemize}
		\item  a power function distribution with density 
		\begin{equation}\label{power}
			g_1(x,\theta)=(\theta+1)x^{\theta},\;x\in(0,1),\;\theta>0;
		\end{equation}
		\item  a distribution with density \begin{equation}\label{g2}
			g_2(x,\theta)=1+\theta(2x-1),\;x\in(0,1),\;\theta>0;
		\end{equation}
		\item a mixture of  a  uniform and  a  power function distributions
		\begin{equation}\label{mix}
			g_3(x,\theta)=1 - \theta + \theta \beta x^{ \beta-1},\;x\in(0,1),\;\theta\in[0,1];
		\end{equation}
		\item a second Ley-Paindaveine alternative (see \cite{ley_paindaveine_2008})  with density function
		\begin{equation}\label{ley}
			g_4(x,\theta)=1-\theta\pi\cos\pi
			x,\;x\in(0,1),\;\theta\in[0,\pi^{-1}].
		\end{equation}
		
	\end{itemize}
	It can be shown that those alternatives belong  to class $\mathcal{G}$. 
	
	In order to calculate local Bahadur efficiencies  of competitors \eqref{kolmogorov}-\eqref{maxcor}, we use the results
	for exact Bahadur slopes from \cite{nikitinKnjiga} and
	\cite{graneTchirina}.

	Since for  the  test with statistic \eqref{hashimoto}, the exact Bahadur slope is not
	derived yet, we do it here.
	Notice that $C_n$ is  a  $U$-statistic with kernel
	$h(X_1,X_2,X_3,X_4)$. It can be easily shown,  that the kernel is weakly
	degenerate. The projection of the kernel on $X_3$ and $X_4$ is equal
	to
	\begin{align*}
		h^{*}(s,t)&=E(h(X_1,X_2,X_3,X_4)|X_3=s,X_4=t)=\frac{s^2t+t^2s}{6}\\&+\frac{\min(s,t)}{18}-\frac{2st}{9}-\frac{s^2t^2}{6}.
	\end{align*}
	According to  \cite[Th.4]{metron}, we have to  find  the smallest
	$\lambda$  that satisfies the integral equation
	\begin{equation*}
		x(s)=\lambda\int\limits_{0}^1h^{*}(s,t)x(t)dt,\;
		\int\limits_{0}^1x(t)dt=0,
	\end{equation*}
	or equivalently
	\begin{equation*}
		x'''(s)=-\frac{\lambda}{18}x'(s),\;x(0)=x(1)=0,\;\int\limits_{0}^1x(t)dt=0.
	\end{equation*}
	This differential equation has a solution if and only if
	$\lambda$ is the solution of the following equation
	\begin{equation*}
		\sin\frac{\sqrt{\lambda}}{
			6 \sqrt{2}} \Big(\sqrt{\lambda} \cos\frac{\sqrt{\lambda}}{6 \sqrt{2}} -
		6 \sqrt{2} \sin\frac{\sqrt{\lambda}}{6 \sqrt{2}}\Big)=0.
	\end{equation*}
	The smallest positive solution   is $\lambda_0=72\pi^2$.

	Therefore, the local Bahadur slope for alternatives from $\mathcal{G}$ is
	\begin{equation*}
		c_{C}(\theta)=\int_{0}^1\int_{0}^1h^*(s,t)h(s)h(t)dsdt\cdot \theta^2+o(\theta^2),\;\;\theta\to 0.
	\end{equation*}
	
	The limit in probability under  a  close alternative from $\mathcal{G}$,  for  all considered statistics can be obtained applying the results from \cite{metron}.
	
	\medskip
	
	In Table \ref{fig: LBE} we present the  local Bahadur efficiencies
	of considered tests for alternatives \eqref{power}-\eqref{ley}.  The efficiencies of our test are comparable with those of classical tests, and  rather high, in comparison to the recent characterization based tests  $Q^{c}$ and $C$. 
	
	\begin{table}[!hhh]\centering
		\caption{Local Bahadur efficiency}
		\bigskip
		\begin{tabular}{ccccccc}
			\hline
			Alternative &$A^2$&$D$  &$T^{(1)}$ &$T^{(2)}$&$Q^{c}$&$C$\\
			\hline
			$g_1$       &0.80 &0.54 &0.73      &0.81     &0.14  &0.37 \\
			$g_2$       &1    &0.75 &0.99      &0.95     &0     &0.66 \\
			$g_3(3)$    &0.96 &0.74 &0.94      &0.82     &0.06  &0.63 \\
			$g_4$       &0.99 &0.81 &1         &0.96     &0     &0.76  \\
		\end{tabular}
		\label{fig: LBE}
	\end{table}
	
	\bigskip

	Although the efficiencies for the considered alternatives are rather high, there exists alternatives for which our tests have even better efficiencies.  Namely, using the  result  from \cite[Th. 6]{metron}, we obtain locally optimal alternatives, i.e. alternatives for which our tests have local Bahadur efficiency 1.

	For example, for the test $T^{(1)}_n$ some of locally optimal alternative densities are
	\begin{align}
		g_{(1),1}(x,\theta)&=1+\theta \cos\pi x,\;\;x\in [0,1],\\
		g_{(1),2}(x,\theta)&=\frac{\sqrt{1-\theta^2}}{1-\theta \cos\pi x}, \;\;x\in [0,1].
	\end{align}

\begin{figure}[h!]
		\begin{center}
			\includegraphics[scale=0.6]{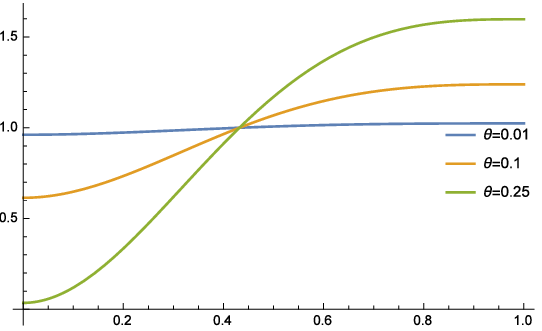}\caption{Locally optimal alternatives $g_{(2),1}(x,\theta)$ }
			\label{fig: optimalne1}
		\end{center}
	\end{figure}

\begin{figure}[h!]
		\begin{center}
			\includegraphics[scale=0.6]{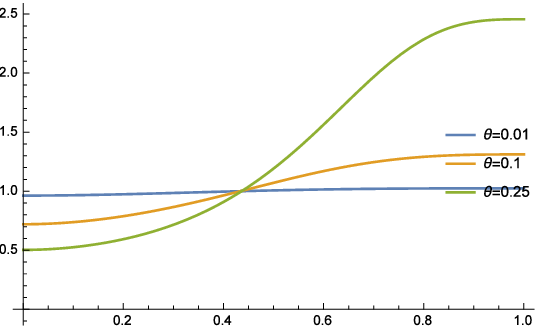}\caption{Locally optimal alternatives  $g_{(2),2}(x,\theta)$ }
			\label{fig: optimalne2}
		\end{center}
	\end{figure}

	The locally optimal densities for $T^{(2)}_n$ are too complicated
	to display.  We present some of them  in Figure \ref{fig: optimalne1} and Figure
	\ref{fig: optimalne2}.


	\subsection{Testing arbitrary simple hypothesis}
	
	\bigskip
	
	We now pass  to  the  testing   of the  null hypothesis that the sample is from  a 
	continuous distribution function $F(x)$,  such that $0<F(x)<1$ for
	all $x\in R$, against   an  alternative  distribution function
	$F(x,\theta)=F(x-\theta),\;\theta>0$.
	
	Denote the alternative
	density function with $f(x,\theta)$ and let
	$h^{\ast}(x)=f'_{\theta}(x,0)$. If  null hypothesis is true, then
	$F(X)$ has  the  uniform $U[0,1]$ distribution and the alternative that
	corresponds to $F(x-\theta)$ is
	\begin{equation*}
		G(x,\theta)=P_{\theta}\{F(X)\leq x\}=P\{X\leq
		F^{-1}(x)\}=F(F^{-1}(x)-\theta),\;x\in(0,1).
	\end{equation*}
	Therefore, $G(x,\theta)$ is   a  close alternative to  the  uniform
	distribution, and $G(x,0)$ is obviously uniform. Then, our testing problem  can
	be reformulated as
	$H_0:$ $U[0,1] (\theta=0)$ against $H_1:$  $G(x,\theta)\;(\theta>0)$.
	The alternative density function is
	\begin{equation*}
		g(x,\theta)=\frac{\partial G(x,\theta)}{\partial
			x}=\frac{f(F^{-1}(x),\theta)}{f(F^{-1}(x))}=\frac{f(F^{-1}(x)-\theta)}{f(F^{-1}(x))}.
	\end{equation*}
	Then we have
	\begin{equation*}
		h(x)=g_{\theta}'(x,0)=\frac{ h^{\ast}(F^{-1}(x))}{f(F^{-1}(x))}.
	\end{equation*}
	Now, the integral in Theorem \ref{slopeNasa}  can be expressed as
	\begin{equation*}
		\begin{aligned}
			&\int_{0}^1\int_{0}^1h(x)h(y)\phi^{\ast}_{m}(x,y)dxdy=\int_{0}^1\int_{0}^1\frac{
				h^{\ast}(F^{-1}(x))}{f(F^{-1}(x))}\frac{
				h^{\ast}(F^{-1}(y))}{f(F^{-1}(y))}\phi^{\ast}_{m}(x,y)dxdy\\
			&=\int_{-\infty}^{\infty}\int_{-\infty}^{\infty}\frac{
				h^{\ast}(u)}{f(u)}\frac{
				h^{\ast}(v)}{f(v)}\phi^{\ast}_{m}(F(u),F(v))f(u)f(v)dudv\\&=\int_{-\infty}^{\infty}\int_{-\infty}^{\infty}
			h^{\ast}(u) h^{\ast}(v)\phi^{\ast}_{m}(F(u),F(v))dudv.
		\end{aligned}
	\end{equation*}
	Since $G(x,\theta)$ and $G(x,0)$ are defined on the same support
	$(0,1)$ we can calculate the Kullback-Leibler distance in terms of
	the Fisher information function which here is  equal to
	\begin{equation*}
		\begin{aligned}
			I&=\int_{0}^1h^{2}(x)dx=\int_{0}^1\big(\frac{
				h^{\ast}(F^{-1}(x))}{f(F^{-1}(x))}\big)^2dx=\int_{-\infty}^{\infty}\big(\frac{
				h^{\ast}(u)}{f(u)}\big)^2f(u)du
			\\ &=\int_{-\infty}^{\infty}\frac{ (h^{\ast}(u))^2}{f(u)}du.
		\end{aligned}
	\end{equation*}
	Therefore, we have all   the  ingredients to calculate  the  local Bahadur efficiencies of  the  proposed tests, in  general  null hypothesis case.
	
	In Table  \ref{fig: LBElocation}, we present the local Bahadur efficiency of our tests applied to standard normal, Cauchy, and logistic null distribution  and corresponding location alternatives. We use notation from \cite[Ch. 2]{nikitinKnjiga}.
	
	\begin{table}[!hhh]\centering
		\caption{Local Bahadur efficiency for location alternatives}
		\bigskip
		\begin{tabular}{lccc}
			\hline
			Statistics &Gaussian&Cauchy&Logistic\\
			\hline
			$A^2$&0.96&0.66&1\\
			$D$&0.64&0.81&0.75\\
			$T^{(1)}$&0.91&0.76&0.99\\
			$T^{(2)}$&0.87&0.72&0.95\\
			$Q^{c}$&0&0&0\\
			$C$&0.49&1&0.66\\
			
		\end{tabular}
		\label{fig: LBElocation}
	\end{table}
	
	It is interesting to note that test \eqref{hashimoto} is locally
	optimal for  the  location alternative of Cauchy distribution.

	\subsection{Testing for uniformity on unknown support}
	In this section we show how we can adapt  the  presented tests to testing  uniformity  on  an  unknown support, i.e for testing $H_0$ that   a  sample $X_1,...,X_n$ comes from  the  uniform $\mathcal{U}[a,b]$ where $a$ and $b$ are unknown.
	
	First, consider  the  transformed ordered sample $Y_2,...,Y_{n-1}$ where \begin{equation*}Y_i=\frac{X_{(i),n}-\hat{a}}{\hat{b}-\hat{a}},\;\;i=2,3,...,n-1\end{equation*}
	and  $\hat{a}=X_{(1),n}$ and $\hat{b}=X_{(n),n}$ are MLE of a and b, respectively.
	
	Each of  the  previously presented tests  can be applied to  the  transformed sample. Following this procedure, we obtain the  tests for rectangularity on  an  unknown support. It is obvious that, under  the  null hypothesis, the distribution of test statistics does not depend on $a$ and $b$. Additionally, our statistics are  symmetric functions of the sample, i.e. we may consider them as  functions of order statistics.
	Moreover, it can be shown   that  random vector $(Y_2,...,Y_{n-1})$ is equally distributed as the vector of order statistics of i.i.d. sample  of size $n-2$ from  uniform $\mathcal{U}[0,1]$.

	Having this in mind,  we conclude that all derived asymptotic properties of  the  proposed tests still hold. 
	
	\section{Power study and applications}\label{sec:power}
	
\begin{figure}[h!]
		\begin{center}
			\includegraphics[scale=0.74]{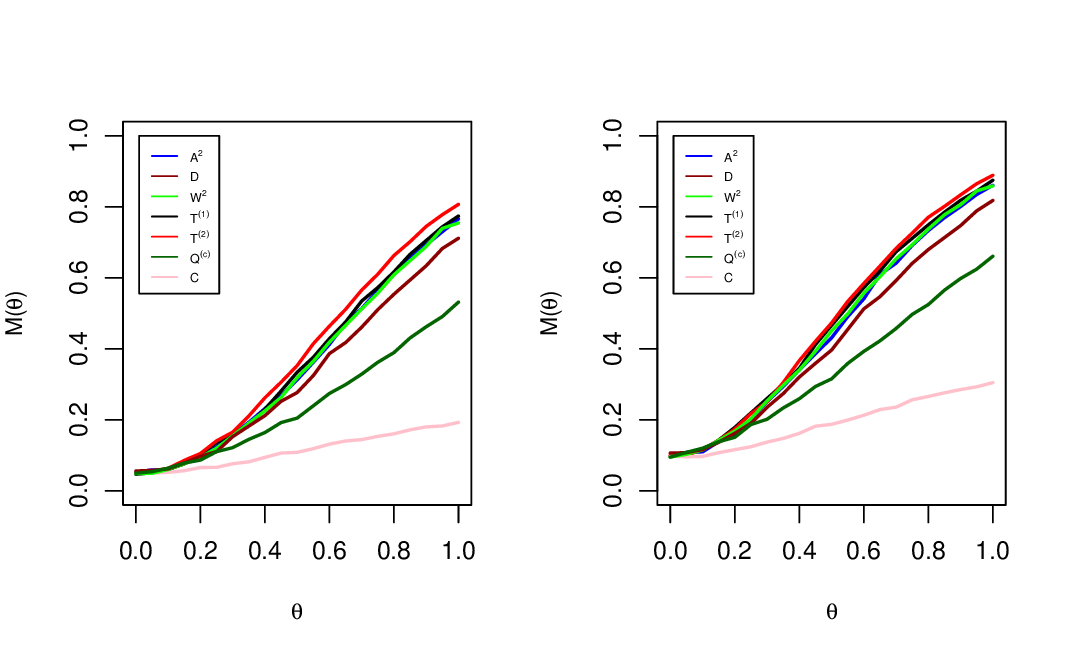}\caption{ Empirical powers of tests for power alternatives for $\alpha=0.05$ (left) and $\alpha=0.1$ (right), $n=20$}
			\label{fig: mocibeta20}
		\end{center}
	\end{figure}

\begin{figure}[h!]
		\begin{center}
			\includegraphics[scale=0.74]{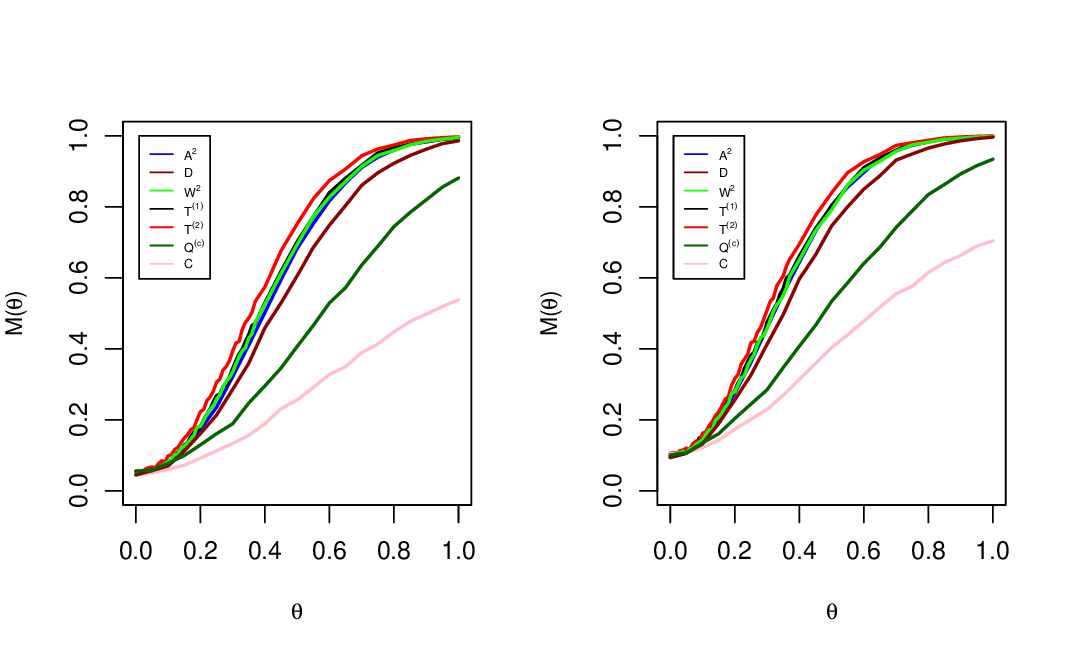}\caption{ Empirical powers of tests for power alternatives for $\alpha=0.05$ (left) and $\alpha=0.1$ (right), $n=50$}
			\label{fig: mocibeta50}
		\end{center}
	\end{figure}

\begin{figure}[h!]
		\begin{center}
			\includegraphics[scale=0.74]{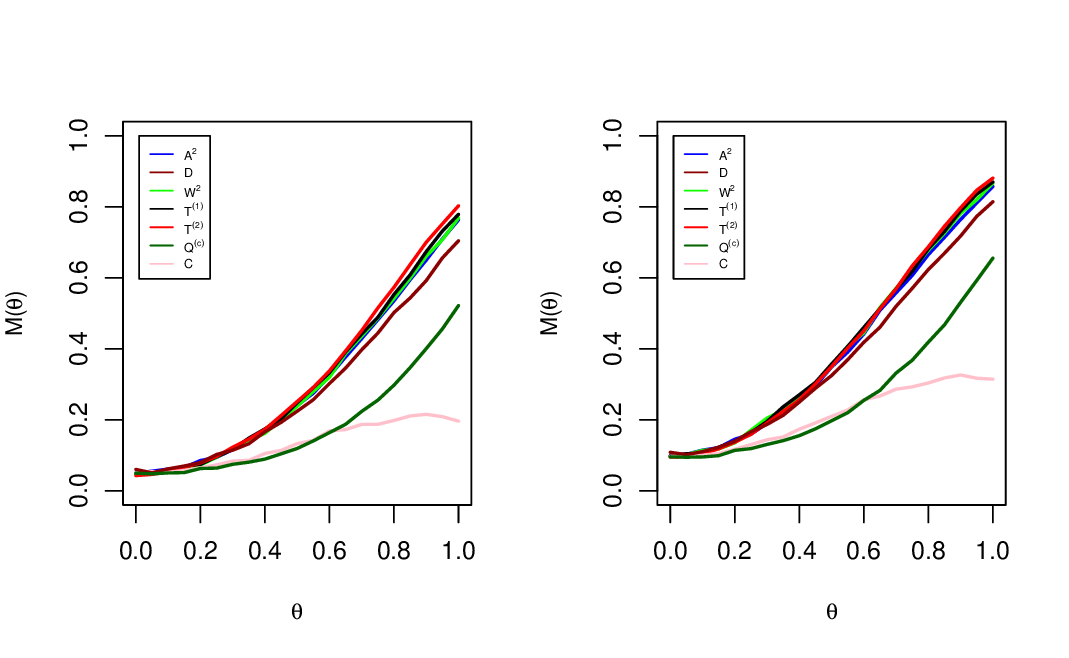}\caption{ Empirical powers of tests for $g_2$ alternatives for $\alpha=0.05$ (left) and $\alpha=0.1$ (right), $n=20$ }
			\label{fig: mocig220}
		\end{center}
	\end{figure}

\begin{figure}[h!]
		\begin{center}
			\includegraphics[scale=0.74]{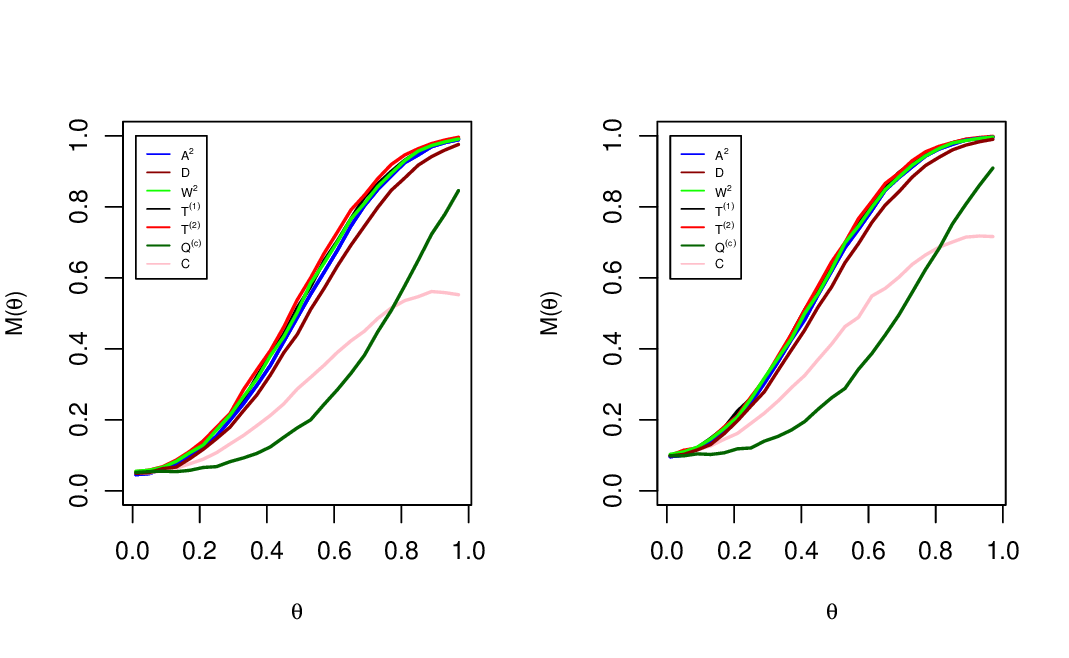}\caption{ Empirical powers of tests for $g_2$ alternatives for $\alpha=0.05$ (left) and $\alpha=0.1$ (right), $n=50$ }
			\label{fig: mocig250}
		\end{center}
	\end{figure}

\begin{figure}[h!]
		\begin{center}
			\includegraphics[scale=0.74]{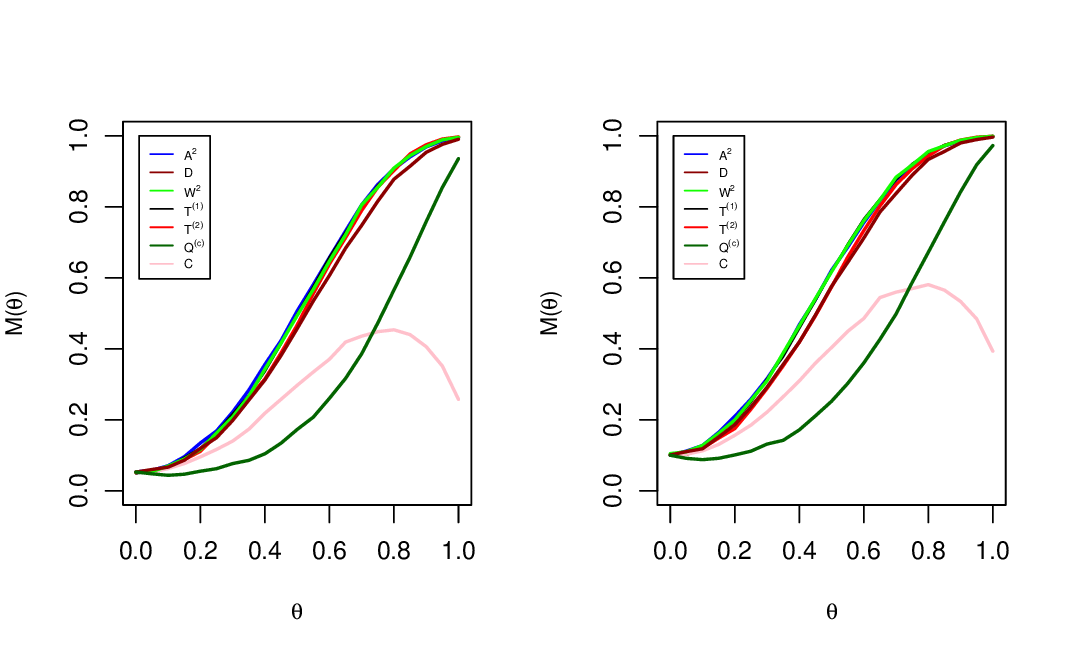}\caption{ Empirical powers of test for mixture with power alternatives for $\alpha=0.05$ (left) and $\alpha=0.1$ (right), $n=20$ }
			\label{fig: mociMesavina320}
		\end{center}
	\end{figure}

\begin{figure}[h!]
		\begin{center}
			\includegraphics[scale=0.74]{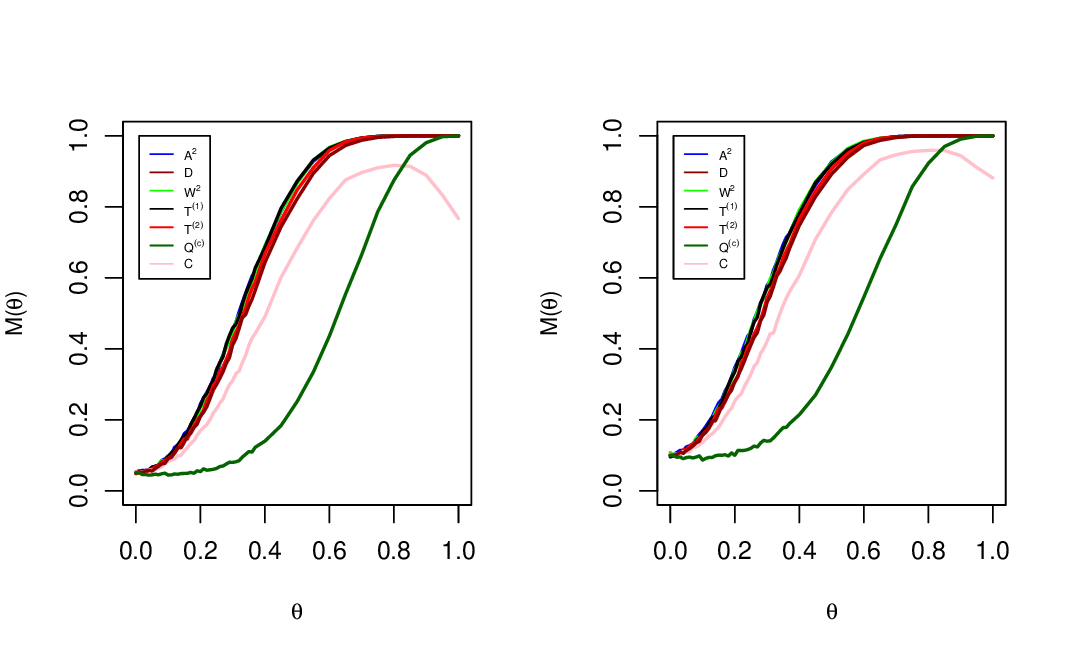}\caption{ Empirical powers of test for mixture with power alternatives for $\alpha=0.05$ (left) and $\alpha=0.1$ (right), $n=50$ }
			\label{fig: mociMesavina350}
		\end{center}
	\end{figure}

\begin{figure}[h!]
		\begin{center}
			\includegraphics[scale=0.74]{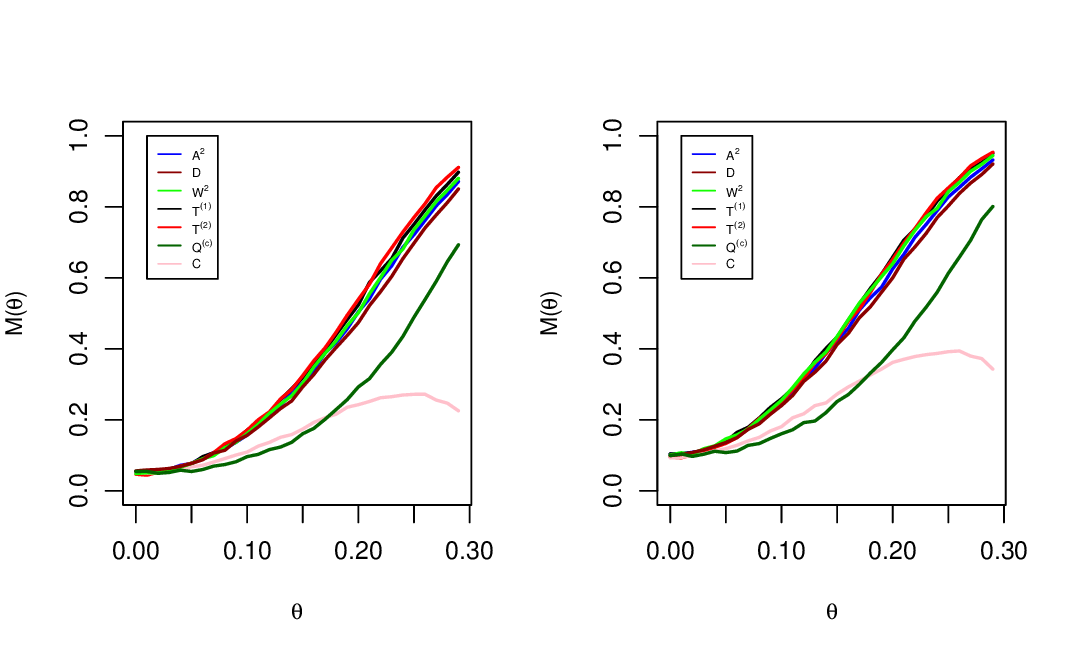}\caption{ Empirical powers of tests for Ley-Paindaveine alternatives  for $\alpha=0.05$ (left) and $\alpha=0.1$ (right), $n=20$ }
			\label{fig: mociLp20}
		\end{center}
	\end{figure}

\begin{figure}[h!]
		\begin{center}
			\includegraphics[scale=0.74]{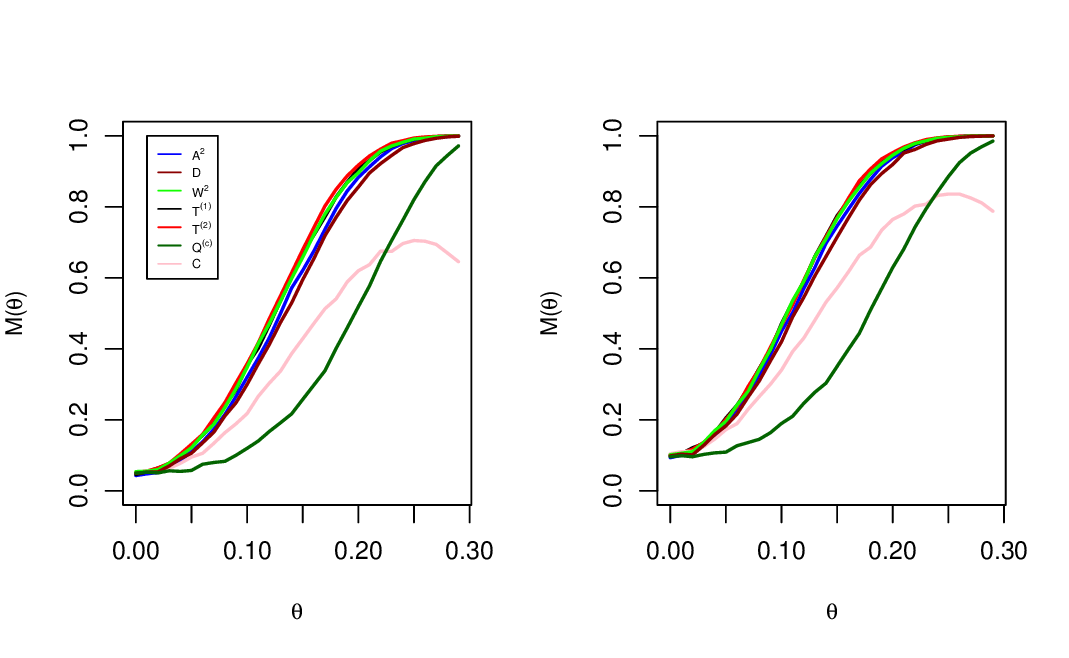}\caption{ Empirical powers of tests for Ley-Paindaveine alternatives  for $\alpha=0.05$ (left) and $\alpha=0.1$ (right), $n=50$ }
			\label{fig: mociLp50}
		\end{center}
	\end{figure}

	The powers of our and competitors tests are presented in Figures
	\ref{fig: mocibeta20}-\ref{fig: mociLp50}. The powers are
	calculated using Monte Carlo simulations with 10000 replicates  at
	levels of significance $0.05$ and $0.1$ for samples of size  $n=20$ and  $n=50$.
	

	From these figures we may conclude
	that for all considered  values of parameter $\theta$, our tests outperform the competitor ones, while  the   test based on  the  maximal covariance characterization is the worst.  In particular, in the case of power and $g_2$ alternatives, $T^{(2)}_n$ is the most powerful one, while in case of other two alternatives  $T^{(2)}_n$ and $W^2_n$ are the leading ones. Also, it is noticeable that  the differences between the estimated powers of the consider tests are larger for smaller level of significance.  
	
	\medskip

	In order to additionally explore small sample properties of the  proposed tests, we also consider the following multi-parameter alternatives defined on [0,1]:
	\begin{itemize}
		\item a beta distribution $\mathcal{B}(\alpha,\beta)$ with density
		\begin{equation*}
			g(x;\alpha, \beta)=\frac{x^{\alpha-1}(1-x)^{\beta-1}}{B(\alpha,\beta)},\;\;\alpha>0,\beta>0;
		\end{equation*}
		\item  a Tukey's distribution $T(a,\lambda)$ (see \cite{joiner1971some}) with inverse d.f.
		\begin{equation*}
			G^{-1}(x;a,\lambda)=\frac{ax^{\lambda}-(1-x)^{\lambda}+1}{a+1},\;\;\;a>0,\lambda>0;
		\end{equation*}
		\item a Johnson's bounded distribution ${JB}(\frac{\gamma}{\delta^2},\frac{1}{\delta^2})$ (see \cite{bowman1983johnson})
		with inverse d.f.
		\begin{equation*}
			G^{-1}(x;\gamma,\delta)=\frac{1}{1+e^{-\frac{\Phi^{-1}(x)-\gamma}{\delta}}},	\;\;\; \gamma \in \mathbf{R}, \delta>0,
		\end{equation*}
		where $\Phi(x)$ is d.f. of standard normal distribution;
		\item a truncated normal distribution $TN(\mu,\sigma^2)$ with density
		\begin{equation*}
			g(x;\mu,\sigma)=\frac{1}{(\Phi(\frac{1-\mu}{\sigma})-\Phi(\frac{-\mu}{\sigma}))\sqrt{2\pi\sigma^2}}e^{-\frac{(x-\mu)^2}{2\sigma^2}},\;\;\;\mu\in \mathbf{R},\sigma>0.
		\end{equation*}
		
	\end{itemize}
	These families  include both symmetric and positively and negatively skewed  distributions with different shapes.
	One can notice that  $\mathcal{B}(1,1)$ and $T(a,1)$ coincide with $\mathcal{U}[0,1]$, while $JB(\frac{\gamma}{\delta^2},\frac{1}{\delta^2})$ and $TN(\mu,\sigma^2)$ are never uniform for any choice of parameters. The results are presented in Tables \ref{fig: ostali20} and \ref{fig: ostali50}.

	\begin{table}[htbp]
		\centering
		\bigskip
		\caption{ Empirical powers of tests  for $n=20$}
		\bigskip
		\centering
		\begin{tabular}{ccccccccc}
			\hline\noalign{\smallskip}
			Alternative  &$\alpha$ &$A^2$&D&$W^2$&$T^{(1)}$&$T^{(2)}$&$Q^c$&C \\\hline
			\multirow{ 2}{*}{$\mathcal{B}(1,2)$}  &0.05 &0.76&0.71&0.76&0.78&0.65&0.53&0.19\\
			&0.1&0.86&0.82&0.86&0.87&0.79&0.66&0.30\\
			\multirow{2}{*}{$\mathcal{B}(1.2,2)$}&0.05&0.49&0.48&0.51&0.54&0.33&0.49&0.11\\
			&0.1&0.64&0.62&0.67&0.69&0.51&0.64&0.19\\
			\multirow{2}{*}{$\mathcal{B}(2,1.2)$}&0.05&0.49&0.48&0.51&0.54&0.61&0.49&0.11\\
			&0.1&0.65&0.63&0.67&0.70&0.74&0.64&0.19\\                              \multirow{ 2}{*}{$\mathcal{B}(1.2,0.8)$}& 0.05&0.34&0.29&0.32&0.33&0.30&0.09&0.15\\
			&0.1&0.46&0.40&0.45&0.45&0.41&0.15&0.23\\ 
			\multirow{ 2}{*}{$\mathcal{B}(3,2)$} &0.05&0.47&0.53&0.49&0.57&0.71&0.89&0.04\\
			&0.1&0.68&0.68&0.70&0.76&0.85&0.95&0.10\\
			\multirow{ 2}{*}{$\mathcal{B}(0.5,0.5)$}  &0.05 &0.54&0.21&0.19&0.16&0.16&0.24&0.21\\
			&0.1&0.66&0.33&0.33&0.27&0.26&0.40&0.28\\
			\multirow{ 2}{*}{$T(2,3)$}            &0.05 &0.23&0.26&0.25&0.27&0.18&0.24&0.17\\
			&0.1&0.35&0.38&0.38&0.40&0.32&0.34&0.27\\
			\multirow{ 2}{*}{$T(3,2)$}           &0.05 &0.28&0.23&0.24&0.24&0.23&0.07&0.16\\
			&0.1&0.38&0.33&0.35&0.34&0.36&0.12&0.24\\
			\multirow{ 2}{*}{$T(0.5,3.5)$}     &0.05  &0.29&0.36&0.33&0.36&0.41&0.37&0.23\\
			&0.1  &0.43&0.47&0.48&0.51&0.53&0.49&0.35\\
			\multirow{ 2}{*}{$T(1,10)$}        &0.05 &0.85&0.85&0.87&0.92&0.82&0.99&0.71\\
			&0.1&0.95&0.94&0.97&0.98&0.92&0.99&0.86\\
			\multirow{ 2}{*}{$JB(0.5,0.5)$}      &0.05  &0.69&0.60&0.63&0.62&0.65&0.10&0.45\\
			&0.1&0.78&0.70&0.73&0.72&0.75&0.15&0.55\\
			\multirow{ 2}{*}{$JB(-0.5,0.5)$}   &0.05 &0.69&0.60&0.62&0.62&0.54&0.10&0.44\\
			&0.1&0.78&0.71&0.73&0.72&0.63&0.15&0.53\\
			\multirow{ 2}{*}{$JB(0.5,1)$} &0.05 &0.46&0.51&0.49&0.57&0.21&0.90&0.05\\
			&0.1&0.67 &0.67 &0.68 &0.75 &0.44 &0.95 &0.11  \\
			\multirow{ 2}{*}{$JB(-0.5,1)$} &0.05 &0.46&0.52&0.49&0.59&0.69&0.90&0.05\\
			&0.1&0.67 &0.67 &0.69 &0.76 &0.83 &0.95 &0.12  \\	
			\multirow{ 2}{*}{$JB(0,0.5)$}&0.05&0.11&0.10&0.09&0.07&0.07&0.02&0.10\\
			&0.1&0.18&0.16&0.15&0.13&0.13&0.06&0.15\\
			\multirow{ 2}{*}{$TN(1,0.5^2)$}&0.05&0.58&0.56&0.58&0.62&0.63&0.36&0.22\\
			&0.1&0.71&0.67&0.71&0.73&0.74&0.48&0.34\\
			\multirow{ 2}{*}{$TN(1.2,0.5^2)$}&0.05&0.85&0.82&0.86&0.87&0.88&0.57&0.34\\
			&0.1&0.92&0.90&0.93&0.93&0.93&0.69&0.47\\ 
			\multirow{ 2}{*}{$TN(0.5,0.2^2)$}&0.05&0.19&0.20&0.16&0.28&0.24&0.94&0.04\\
			&0.1&0.45&0.39&0.42&0.56&0.47&0.97&0.11\\                                          
		\end{tabular}
		\label{fig: ostali20}
	\end{table}
	
	\begin{table}[htbp]
		\centering
		\bigskip
		\caption{ Empirical powers of tests  for $n=50$}
		\bigskip
		\centering
		\begin{tabular}{ccccccccc}
			\hline\noalign{\smallskip}
			Alternative  &$\alpha$ &$A^2$&D&$W^2$&$T^{(1)}$&$T^{(2)}$&$Q^c$&C \\\hline
			\multirow{ 2}{*}{$\mathcal{B}(1,2)$}  &0.05 &0.99&0.98&1&1&0.98&0.88&0.53\\
			&0.1&1&0.99&1&1&0.99&0.94&0.71\\
			\multirow{2}{*}{$\mathcal{B}(1.2,2)$}&0.05&0.94&0.90&0.94&0.94&0.81&0.87&0.32\\
			&0.1&0.98&0.95&0.97&0.97&0.92&0.93&0.47\\
			\multirow{2}{*}{$\mathcal{B}(2,1.2)$}&0.05&0.94&0.91&0.93&0.94&0.97&0.87&0.32\\
			&0.1&0.97&0.95&0.97&0.97&0.99&0.93&0.47\\                              \multirow{ 2}{*}{$\mathcal{B}(1.2,0.8)$}& 0.05&0.70&0.58&0.66&0.66&0.63&0.12&0.35\\
			&0.1&0.79&0.70&0.77&0.76&0.74&0.19&0.46\\ 
			\multirow{ 2}{*}{$\mathcal{B}(3,2)$} &0.05&0.98&0.96&0.98&0.99&1&1&0.15\\
			&0.1&1&0.99&0.99&1&1&1&0.30\\
			\multirow{ 2}{*}{$\mathcal{B}(0.5,0.5)$}  &0.05 &0.87&0.43&0.45&0.40&0.40&0.84&0.42\\
			&0.1&0.94&0.60&0.66&0.58&0.55&0.91&0.54\\
			\multirow{ 2}{*}{$T(2,3)$}            &0.05 &0.55&0.57&0.59&0.61&0.49&0.40&0.45\\
			&0.1&0.69&0.69&0.72&0.72&0.66&0.52&0.58\\
			\multirow{ 2}{*}{$T(3,2)$}           &0.05 &0.54&0.46&0.52&0.52&0.54&0.07&0.35\\
			&0.1&0.66&0.58&0.64&0.64&0.66&0.13&0.46\\
			\multirow{ 2}{*}{$T(0.5,3.5)$}     &0.05  &0.71&0.74&0.75&0.76&0.80&0.66&0.63\\
			&0.1  &0.83&0.83&0.86&0.86&0.88&0.76&0.75\\
			\multirow{ 2}{*}{$T(1,10)$}        &0.05 &1&1&1&1&1&1&1\\
			&0.1&1&1&1&1&1&1&1\\
			\multirow{ 2}{*}{$JB(0.5,0.5)$}      &0.05  &0.97&0.93&0.95&0.95&0.96&0.11&0.84\\
			&0.1&0.98&0.96&0.97&0.97&0.98&0.17&0.89\\
			\multirow{ 2}{*}{$JB(-0.5,0.5)$}   &0.05 &0.96&0.93&0.95&0.95&0.90&0.11&0.83\\
			&0.1&0.98&0.96&0.98&0.97&0.94&0.17&0.89\\
			\multirow{ 2}{*}{$JB(0.5,1)$} &0.05 &0.98&0.96&0.97&0.98&0.79&1&0.18\\
			&0.1&0.99 &0.98 &0.99 &0.99 &0.94 &1 &0.33  \\
			\multirow{ 2}{*}{$JB(-0.5,1)$} &0.05 &0.99&0.96&0.97&0.98&0.99&1&0.19\\
			&0.1&1 &0.99 &0.99 &1 &1 &1 &0.35  \\	
			\multirow{ 2}{*}{$JB(0,0.5)$}&0.05&0.14&0.11&0.10&0.09&0.09&0.12&0.13\\
			&0.1&0.24&0.20&0.19&0.17&0.18&0.22&0.22\\
			\multirow{ 2}{*}{$TN(1,0.5^2)$}&0.05&0.94&0.92&0.95&0.95&0.96&0.67&0.61\\
			&0.1&0.97&0.97&0.98&0.98&0.98&0.77&0.74\\ 
			\multirow{ 2}{*}{$TN(1.2,0.5^2)$}&0.05&1&1&1&1&1&0.89&0.82\\
			&0.1&1&1&1&1&1&0.94&0.90\\ 
			\multirow{ 2}{*}{$TN(0.5,0.2^2)$}&0.05&0.93&0.75&0.86&0.92&0.84&0.99&0.21\\
			&0.1&0.99&0.90&0.97&0.98&0.96&1&0.44\\                                         
		\end{tabular}
		\label{fig: ostali50}
	\end{table}
	
	Taking into account the previous results, and those presented in Tables \ref{fig: ostali20} and \ref{fig: ostali50}, we  may conclude that, although asymptotically equivalent, the tests $T_n^{(1)}$ and $W^2_n$ do differ for small sample size $n=20$. Moreover, in that case, $T_n^{(1)}$ is usually more powerful than  $W^2_n$.  
	
	In general, for the sample size $n=20$, the "ordering" of the considered tests largely depends on the shape of the  alternative distribution. 
	In particular, for  alternatives with monotone densities, the classical tests $A^2_n, D_n,W^2_n$,  and our two tests $T^{(1)}_n$ and  $T^{(2)}_n$ are shown to be much more powerful than $Q^{c}_n$ and $C.$ Usually $T^{(2)}_n$ is the leading one.  
	Also, $T^{(1)})n$ and $T^{(2)}_n$ are among the best, for positively and negatively skewed alternatives, respectively, and the only test that is sometimes better in these cases is $Q^c_n$. In the case of U-shaped  densities, $A^2_n$ is the leading one.  For the grater sample size $n=50$, the "ordering" of the tests is very similar.
	

	
	\bigskip
	
	\subsection{Specification tests for conditional duration models}
	
	Recently, high-frequency data have become widely available in markets. Therefore, there is a growing interest in modeling such data with special attention  given to  modeling the times between observations.      \cite{engle1988} and   \cite{dufour2000},  using the idea from GARCH,
	proposed to model  durations (appropriately standardized) with 
	\begin{equation*} \label{duration} x_t=\mu_t \varepsilon_t, \ \mu_t=a_0+\sum_{j=1}^p a_j x_{t-j}+\sum_{j=1}^q b_j\tilde \mu_{t-j},
	\end{equation*}  where $\varepsilon_t$ is a sequence of i.i.d. positive random variables with d.f. F and  the  expectation one, so called \emph{innovation process}.  This model is widely known as \emph{Autoregressive Conditional Duration Model} and labeled with ACD(p,q). 
	Naturally, developing modeling diagnostic tools have become important. One of two possible directions  is  to inspect the adequacy of  the  functional form of   the  conditional duration, while the other is to test the distribution of  the  error term. 
	
	In this section we compare  the  introduced tests as the specification tests for ACD model with exponential innovations ($\rm EACD$ model).  Since the innovations are unobservable, it is natural to base a test on corresponding residuals given by
	\begin{equation*}
		\widehat{\varepsilon}_t=\frac{x_t}{\widehat{\mu}_t},\; t=1,...,T,
	\end{equation*}
	where $\widehat{\mu}_t$ is the estimator of conditional mean duration.
	Taking into account the model definition, the null hypothesis is that the corresponding innovations have exponential $\mathcal{E}(1)$ distribution. Hence, we apply our tests to  the   appropriately transformed  residuals. Following the procedure described in \cite{meintanisACD}, we obtain  the   empirical powers of considered tests. In order to use results from \cite{meintanisACD} as  benchmark  ones, we  consider  $ACD(1,1)$ model with parameters $a_0=0.3, a_1=0.3$ and $b_1=0.4$ and sample sizes $T=100$ and $T=200$. For alternative distribution we choose Weibull, gamma and lognormal distributions (see \cite{meintanisACD}).
	
	From  Tables \ref{fig: exp100} and \ref{fig: exp200} and the results presented in \cite{meintanisACD}, we can notice that all empirical sizes are satisfactory and that, in general, the powers are reasonably high for all tests. In particular, in  the  case of gamma and Weibull alternative, tests based on Too-Lin characterization are among the best. They are more powerful than $C_n$, $Q^c_n$,  and classical $D_n$ tests and comparable with $A^2_n$ and $W^2_n$ tests.  A  similar situation is in the case of lognormal alternative with the exception of $C_n$ that  outperforms  others for shape parameter $\theta=1.1$.

	\begin{table}[htbp]
		\centering
		\bigskip
		\caption{  Empirical powers of tests for exponential innovations,  for $T=100$}
		\bigskip
		\centering
		\begin{tabular}{cccccccc}
			\hline\noalign{\smallskip}
			Alternative   &$A^2$&D&$W^2$&$T^{(1)}$&$T^{(2)}$&$Q^C$&C \\\hline
			E(1) &0.06&0.06&0.06&0.05&0.06&0.05&0.04\\\hline
			W(1.1)&0.24&0.18&0.20&0.22&0.23&0.21&0.10\\
			W(1.2)&0.56&0.46&0.54&0.57&0.56&0.57&0.19\\
			W(1.3)&0.88&0.78&0.85&0.87&0.87&0.83&0.33\\
			W(1.4)&0.98&0.95&0.98&1&1&0.98&0.50\\
			W(1.5)&1&&1&1&1&1&0.67\\
			G(1.2)&0.25&0.20&0.24&0.23&0.25&9.30&0.10 \\
			G(1.3)&0.44&0.39&0.42&0.46&0.48&0.51&0.13 \\
			G(1.4)&0.70&0.52&0.65&0.63&0.65&0.71&0.17\\
			G(1.5)&0.81&0.71&0.78&0.78&0.81&0.84&0.23\\
			LN(0.8)&1&0.97&0.98&0.99&1&1&0.27\\
			LN(0.9)&0.89&0.69&0.73&0.81&0.87&0.3&0.34\\
			LN(1)&0.65&0.41&0.49&0.51&0.55&0.62&0.46\\
			LN(1.1)&0.58&0.49&0.54&0.55&0.51&0.20&0.65\\\hline
		\end{tabular}
		\label{fig: exp100}
	\end{table}
	
	\begin{table}[htbp]
		\centering
		\bigskip
		\caption{ Empirical powers of tests for exponential innovations,  for $T=200$}
		\bigskip
		\centering
		\begin{tabular}{cccccccc}
			\hline\noalign{\smallskip}
			Alternative   &$A^2$&D&$W^2$&$T^{(1)}$&$T^{(2)}$&$Q^c$&C \\\hline
			E(1) &0.05&0.06&0.05&0.04&0.04&0.04&0.05\\\hline
			W(1.1)&0.33&0.31&0.33&0.36&0.37&0.35&0.15\\
			W(1.2)&0.86&0.75&0.85&0.86&0.86&0.84&0.40\\
			W(1.3)&1&0.97&1&1&1&0.99&0.66\\
			W(1.4)&1&1&1&1&1&1&0.86\\
			G(1.2)&0.47&0.34&0.42&0.43&0.47&0.48&0.13 \\
			G(1.3)&0.75&0.62&0.71&0.72&0.76&0.77&0.21 \\
			G(1.4)&0.96&0.83&0.93&0.90&0.91&0.93&0.32\\
			G(1.5)&0.99&0.94&0.97&0.97&0.98&0.99&0.45\\
			LN(0.8)&1&1&1&1&1&1&0.26\\
			LN(0.9)&1&0.98&0.99&0.99&1&1&0.69\\
			LN(1)&0.98&0.77&0.89&0.92&0.92&0.89&0.81\\
			LN(1.1)&0.95&0.81&0.90&0.91&0.88&0.37&0.93\\\hline
		\end{tabular}
		\label{fig: exp200}
	\end{table}

	\subsection{Detecting hidden periodicity}
	We shall demonstrate the use of  the  cumulative periodogram as a useful diagnostic tool  for hidden periodicity of the unspecified frequencies.  For more on theory of analysis of time series in frequency domain we refer to  \cite{brockwell}.
	
	Let us consider   a  real-valued stationary time series $\{X_t,\;t\geq 0\}$, and let $f(\omega_i)$  and $I(\omega_i)$ be the corresponding spectral density of $X_t$ and its estimate based on the first $n$ elements of $X_t$, evaluated at  Fourier frequencies $\omega_i=2\pi i/n$. 
	In \cite{brockwell} it is proved that $\{X_t,0\leq t\leq n\}$ is Gaussian white noise if and only if the random variables
	\begin{equation*}
		Y_k=\frac{\sum_{i=1}^kI(\omega_i)}{\sum_{i=1}^{q-1}I(\omega_i)},\;\;k=1...q-1,\;\;q=\Big[\frac{n-1}{2}\Big],
	\end{equation*}
	are distributed as the order statistics from uniform $\mathcal{U}[0,1]$ distribution.
	Hence, we can use the  uniformity tests to check whether   $\{X_t\}$ is Gaussian white noise.
	
	We compare the powers of presented tests for artificially generated  time series
	
	\begin{equation}
		X_t=\sin\frac{\pi t}{a}+Z_t,\; t=1...T,
	\end{equation}
	where $\{Z_t\}$ is Gaussian white noise.  The powers are estimated based on $N=10000$ Monte Carlo replicates and for level of significance $\alpha=0.05$, for different values of parameter $a$. The values are given in Tables \ref{tabelaGaus} and \ref{tabelaGaus1}.
	
	\begin{table}[ht]
		\centering
		\begin{tabular}{l|rrrrrrr}
			\hline
			$a$ & $A^2$ & $D$ & $W^2$ & $T^{(1)}$&$T^{(2)}$ &$Q^c$ & $C$ \\ 
			\hline
			$\infty$ & 0.07 & 0.07 & 0.07 & 0.06 & 0.07 & 0.05 & 0.07 \\ 
			1 & 0.07 & 0.06 & 0.06 & 0.06 & 0.06 & 0.05 & 0.07 \\
			2 & 0.63 & 0.67 & 0.57 & 0.49 & 0.50 & 0.65 & 0.91 \\ 
			3 & 0.70 & 0.77 & 0.68 & 0.64 & 0.64 & 0.23 & 0.94 \\ 
			4 & 0.78 & 0.85 & 0.79 & 0.78 & 0.80 & 0.00 & 0.96 \\ 
			5 & 0.85 & 0.91 & 0.87 & 0.86 & 0.90 & 0.02 & 0.97 \\ 
			6 & 0.88 & 0.92 & 0.89 & 0.89 & 0.92 & 0.13 & 0.97 \\ 
			7 & 0.91 & 0.94 & 0.92 & 0.92 & 0.95 & 0.40 & 0.97 \\ 
			8 & 0.91 & 0.95 & 0.93 & 0.92 & 0.96 & 0.61 & 0.97 \\ 
			9 & 0.92 & 0.96 & 0.93 & 0.94 & 0.96 & 0.74 & 0.96 \\ 
			10 & 0.95 & 0.97 & 0.95 & 0.95 & 0.98 & 0.84 & 0.96 \\ 
			\hline
		\end{tabular}
		\caption{Empirical powers  in case of testing for hidden periodicities, for $T=100$}
		\label{tabelaGaus}
	\end{table}
	
	\begin{table}[ht]
		\centering
		\begin{tabular}{rrrrrrrr}
			\hline
			$a$ & $A^2$ & $D$ & $W^2$ & $T^{(1)}$&$T^{(2)}$ &$Q^c$ & $C$ \\  
			\hline
			$\infty$ & 0.06 & 0.07 & 0.07 & 0.07 & 0.06 & 0.06 & 0.07 \\ 
			1 & 0.07 & 0.07 & 0.06 & 0.07 & 0.06 & 0.05 & 0.07 \\ 
			2 & 0.96 & 0.98 & 0.97 & 0.96 & 0.94 & 0.95 & 1.00 \\ 
			3 & 0.97 & 0.99 & 0.97 & 0.97 & 0.98 & 0.65 & 1.00 \\ 
			4 & 0.99 & 0.99 & 0.99 & 0.99 & 0.99 & 0.02 & 1.00 \\ 
			5 & 0.99 & 1.00 & 1.00 & 1.00 & 1.00 & 0.02 & 1.00 \\ 
			6 & 1.00 & 1.00 & 1.00 & 1.00 & 1.00 & 0.25 & 1.00 \\ 
			7 & 1.00 & 1.00 & 1.00 & 1.00 & 1.00 & 0.63 & 1.00 \\ 
			8 & 1.00 & 1.00 & 1.00 & 1.00 & 1.00 & 0.85 & 1.00 \\ 
			9 & 1.00 & 1.00 & 1.00 & 1.00 & 1.00 & 0.94 & 1.00 \\ 
			10 & 1.00 & 1.00 & 1.00 & 1.00 & 1.00 & 0.97 & 1.00 \\ 
			\hline
		\end{tabular}
		\caption{Empirical powers   in case of testing for hidden periodicities, for $T=200$}
		\label{tabelaGaus1}
	\end{table}
	We notice that our tests perform well in all cases, while the test based on maximal correlation is the worst one, and, surprisingly, the  test  based on maximal covariance characterization outperformed   the  others.
	The similar conclusion is for  the  series of size 200. In many cases, the  power  are very high,  and  the "ordering" for each value of $a$ is the same.
	
	\section*{Acknowledgments}
	
	This work was supported by the  Ministry of Education, Science and Technological
	Development of Republic of Serbia under Grant 174012.

	We would like to thank the Referees for their useful remarks.

	\bibliographystyle{plain}
	\bibliography{BibliotekaKarakterizacije}

\end{document}